\documentclass[sigconf, nonacm, 10pt]{acmart}

\setcopyright{none}

\begin{document}

\title{An Elastic Ephemeral Datastore using Cheap, Transient Cloud Resources}

\author{Malte Brodmann$^{\ast}$, Nikolas Ioannou, Bernard Metzler, Jonas Pfefferle}
\affiliation{%
  \institution{IBM Research - Zurich}
  \country{Switzerland}
}

\author{Ana Klimovic}
\affiliation{%
  \institution{ETH Zurich}
  \country{Switzerland}
}

\thanks{$^{\ast}$Now at Google}

\begin{abstract}
Spot instances are virtual machines offered at 60-90\% lower cost  that can be reclaimed at any time, with only a short warning period. Spot instances have already been used to significantly reduce the cost of processing workloads in the cloud. However, leveraging spot instances to reduce the cost of stateful cloud applications is much more challenging, as the sudden preemptions lead to data loss. In this work, we propose leveraging spot instances to decrease the cost of ephemeral data management in distributed data analytics applications. We specifically target ephemeral data as this large class of data in modern analytics workloads has low durability requirements; if lost, the data can be regenerated by re-executing compute tasks. We design an elastic, distributed ephemeral datastore that handles node preemptions transparently to user applications and minimizes data loss by redistributing data during node preemption warning periods. We implement our elastic datastore on top of the Apache Crail datastore and evaluate the system with various workloads and VM types. By leveraging spot instances, we show that we can run TPC-DS queries with 60\% lower cost compared to using on-demand VMs for the datastore, while only increasing end-to-end execution time by 2.1\%.
\end{abstract}



%



\maketitle

\sloppy

\section{Introduction}
Cloud spot instances are virtual machines offered in today's public clouds at up to 90\% lower costs~\cite{aws_spot_instances_website,gce_spot_instances_website,azure_spot_instances_website,ibm_transient_servers_website}. However, cloud providers can reclaim spot instances at any time after a short warning period.
Previous work has leveraged spot instances for reducing service costs in different contexts \cite{wang2017exploiting,xu2016blending,ambati2020providing,chohan2010see,yan2016tr,sharma2016flint}. Due to the challenges arising from sudden spot instance preemptions, most prior work has used the compute and memory resources of spot instances to run stateless services, while relying on external durable storage solutions for managing state.\newline
In this work, we ask the question: \emph{can spot instances be leveraged for stateful services like datastores?}
We show how to use spot instance resources to run elastic, distributed datastore services at significantly lower cost compared to using dedicated VMs. We specifically target ephemeral data, which is intermediate data that is produced and consumed at large scale by modern data analytic engines and serverless applications, such as shuffle data. A key property of ephemeral data is that it can be re-computed in case of data loss, by re-executing application tasks~\cite{zaharia2012resilient}. However, frequent re-computations can significantly affect application performance~\cite{li2014tachyon,yan2016tr}. To minimize the impact of spot instance preemptions, we design an elastic datastore that minimizes data loss upon spot instance preemption by efficiently redistributing data between nodes during the spot instance preemption notice period.\newline
Our key contribution is the design and implementation of mechanisms to efficiently redistribute data from about-to-be-preempted nodes to other nodes in the datastore cluster, while allowing user applications to continue interacting with the datastore uninterrupted. \newline
We evaluate our datastore on Google Cloud spot instances and in a private cluster, where we model spot instance preemptions with a probabilistic model~\cite{kadupitige2020modeling}. When running queries from the TPC-DS benchmark suite we observed only a 2-3\% slowdown in end-to-end execution time while reducing the datastore deployment cost by 60\%. We find that read-intensive workloads with long-lived data experience more slowdowns (up to 46\% for a read-only workload with frequent spot instance preemptions) due to imbalances in the datastore caused by data re-distributions, while write-heavy workloads experience negligible overhead. Our results show that spot instances are an attractive platform for running not only stateless compute jobs, but also an ephemeral, elastic, distributed datastore.

\section{Background}
\textbf{Spot instances:} Datacenters resources are often under-utilized as providers provision resources for peak loads and varying customer demands~\cite{clusterdata:Wilkes2011,clusterdata:Wilkes2020,delimitrou2014quasar,43017}. To monetize idle resources cloud providers offer \emph{spot instances}~\cite{aws_spot_instances_website,gce_spot_instances_website,azure_spot_instances_website,ibm_transient_servers_website}, to sell their excess capacities to customers at significant discounts. However, a provider can always reclaim spot instances at any time when the resources are required again. Before evicting a spot instance, cloud providers usually give a warning period (typically 30 seconds to 2 minutes). Users can take advantage of the termination notice period to perform final actions, such as saving critical state. All data stored either within the memory or the local storage volumes of an instance will be lost after the instance is terminated. \newline
\textbf{Ephemeral data:} Ephemeral data refers to data with a limited lifetime and low durability requirements, such as intermediate results generated and consumed between compute stages in distributed data processing engines like Spark or serverless data analytics frameworks~\cite{klimovic2018understanding,klimovic2018pocket,zaharia2012resilient}. In case of data loss, ephemeral data can be re-generated by re-executing compute tasks. However, frequent data loss leads to expensive re-computations that affect application performance~\cite{li2014tachyon,yan2016tr}. Efficient storage and ingestion of ephemeral data is increasingly critical to the end-to-end performance and cost of modern data processing applications~\cite{shen2020magnet,zhang2018riffle}. \newline
\textbf{Crail distributed, ephemeral datastore:} Apache Crail (incubating) is a high-performance distributed datastore designed for ephemeral data management. Crail's architecture serves as a foundation for our work, on top of which we introduce efficient data re-distribution mechanisms to make the datastore attractive to run on low-cost, spot instances. The Crail architecture consists of namenodes, datanodes, and clients. Namenodes store and maintain a hierarchical namespace along with corresponding metadata in memory. Data objects are stored on datanodes, across one or more fixed-sized blocks depending on the object size. Applications use a client library to interact with Crail by creating, reading, or deleting objects. The library performs metadata and data plane operations on behalf of the client. Crail supports various storage technologies (DRAM, SSD, disk) and network technologies (TCP and RDMA). We use the TCP-based DRAM tier in our prototype evaluation since we target data analytics applications that require high-bandwidth access to the ephemeral data store.

\section{A Datastore using Cloud Spot Instances}
\begin{figure}%
  \centering%
  \includegraphics[width=0.90\linewidth]{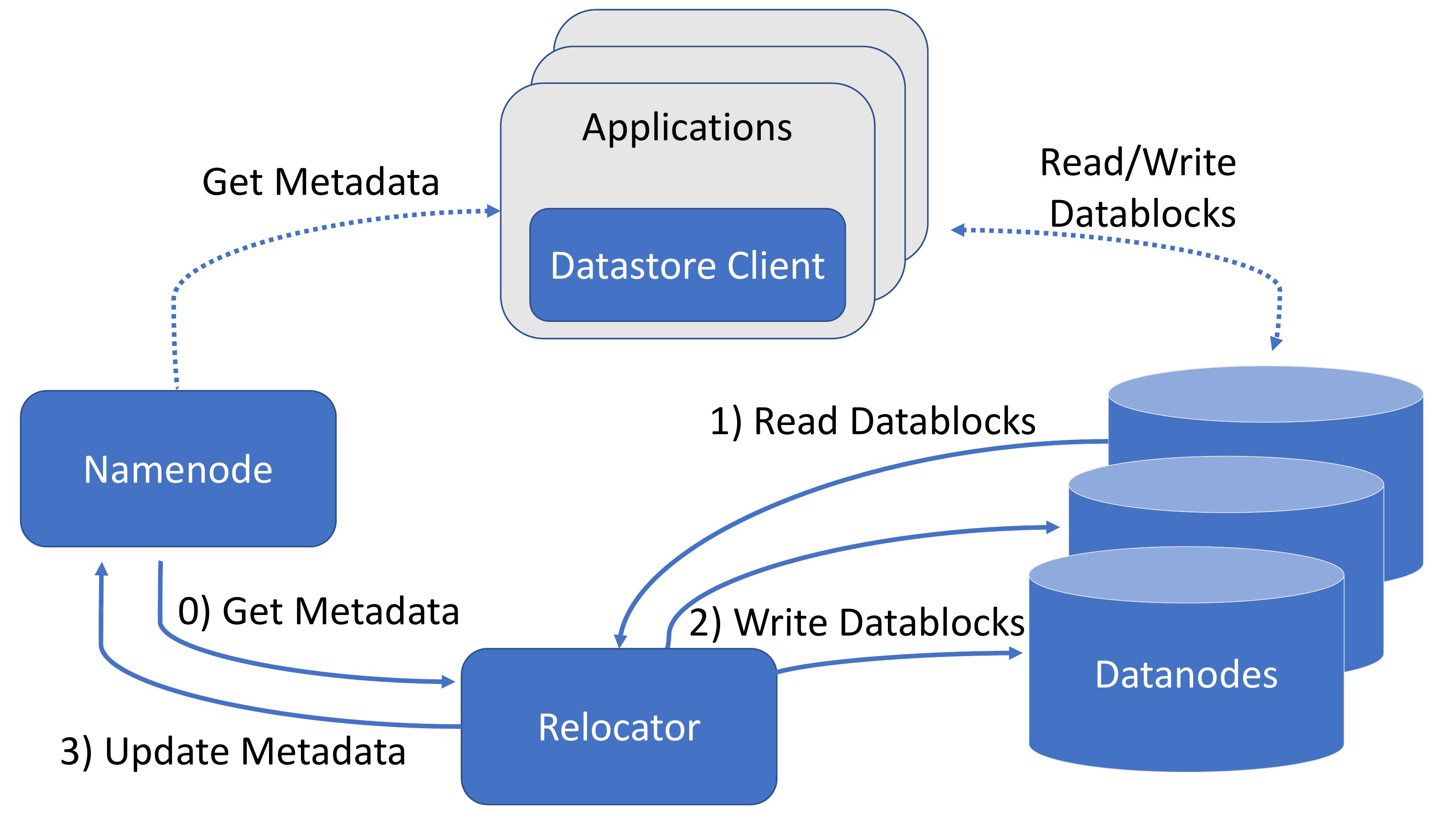}%
  \captionof{figure}{Architectural overview of our redistribution mechanism. Step 0) is performed once to fetch the list of all affected blocks on the respective datanode. Steps 1) -- 3) are repeated for each block.}%
  \label{fig:overview}%
\end{figure}%
To operate a distributed data service on spot instances, we must solve the following challenges: 
\begin{enumerate}
    \item Enable the efficient redistribution of data from one spot instance to other available locations.
    \item Allow applications to continue interacting with the datastore during data redistribution, with minimal performance impact.
    \item Preserve data coherence, during data migration.
\end{enumerate}

\subsection{Data redistribution mechanism}
\label{sec:data_redistribution_mechanism}
When running on traditional VMs, a datastore like Crail can statically map data blocks to datanodes. However, to handle datanode preemptions, we propose to dynamically re-map data blocks across datanodes when a node preemption notice occurs. Our system will begin re-distributing data blocks across available datanodes as soon as a preemption notice is received for a particular datanode until the spot instance is evicted. To minimize data movement, we recommend provisioning  datanodes whose storage capacity is less than or equal to the volume of data that can be moved over the spot VM's network link during the preemption notice period (e.g., 30 seconds, see Figure~\ref{fig:relocation_time}). \newline
Figure~\ref{fig:overview} shows the data relocation process, which we implement by introducing a Relocator service. The Relocator operates independently of the namenode and datanodes. The Relocator performs the following steps for as many blocks as possible from an about-to-be-preempted node to migrate its data:
\begin{enumerate}
  \item Read data from the old block location.
  \item Request a new block location from the namenode. Write the data from the old block location to the new block location.
  \item Notify the namenode to update its internal state to reflect the new block location. Clients requesting this data block will now be directed to its new location.
\end{enumerate}

\subsection{Application transparent data migration}
One challenge with the data redistribution mechanism described in Section~\ref{sec:data_redistribution_mechanism} is that clients often cache block location metadata to optimize performance and avoid frequent metadata lookups for repeated accesses to the same data block. During data relocation, cached block metadata are at risk of becoming stale. This happens when the relocator completes the relocation of a block and updates the metadata by contacting the namenode. Since these updates are not propagated to Crail clients, data operations that use stale metadata will try to contact datanodes that are no longer running.
To allow clients to continue interacting with the datastore during data redistributions, we modify the internals of the client such that failed data operations invalidate cached metadata. Afterwards, the client fetches the most recent metadata and can retry the data operation. \newline
An additional challenge is to ensure the coherence of data during redistributions. In particular, we identified a race condition that occurs when clients write data to a block
that is stored on a datanode whose data is currently being relocated. We prevent incoherent data by rejecting all write operations to datanodes that are in the process of being relocated.
Once the relocation completes, clients can finally execute their write operation at the updated location. Read and delete operations in turn can also be served when a data relocation is taking place.

\begin{figure}%
    \centering%
    \includegraphics[width=0.75\linewidth]{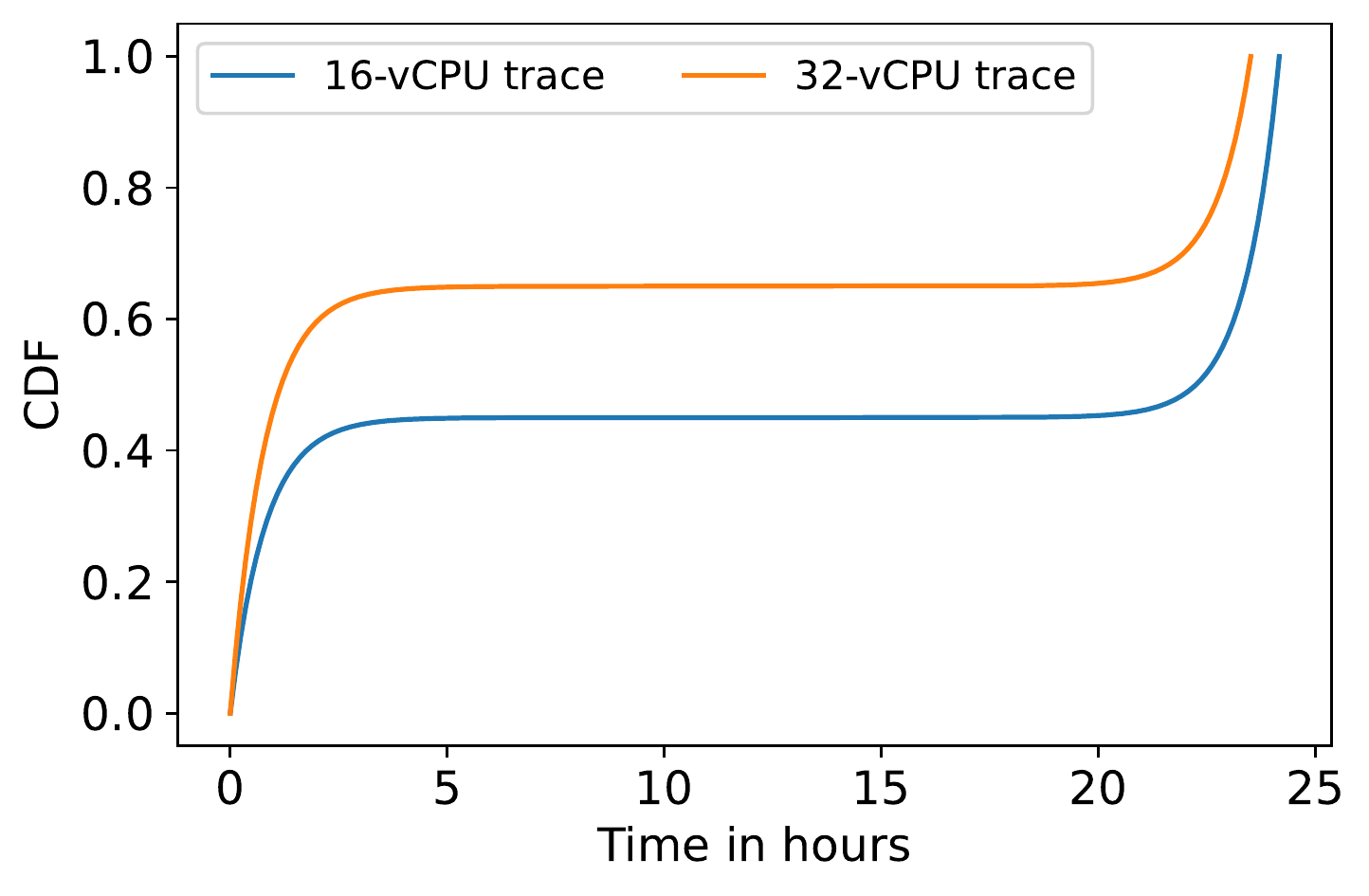}%
    \captionof{figure}{CDFs of spot instance preemption times when using the model parameters of our two generated traces.}%
    \label{fig:cdfs}%
  \end{figure}%

\subsection{Dealing with Data Loss}
In case we are not able to migrate all data blocks from a datanode during the preemption notice period, the remaining data blocks will be lost when the spot instance is terminated. 
Crail -- as well as our extensions to the datastore -- does not rely on data replication for fault tolerance, since this redundancy contributes to significant extra storage costs and
increases complexity as multiple copies of each block would need to be kept consistent.
Instead, we rely on the fact that ephemeral data can be recomputed by the application framework by re-executing compute tasks. Many analytics frameworks,
such as Spark~\cite{zaharia2012resilient}, track data lineage and hence can re-execute the appropriate tasks when the ephemeral datastore returns a `data unavailable' error.

\subsection{Limitations}
Our design does not support the preemption of name\-nodes as Crail's architecture does not support moving or replicating metadata without substantial rework. Therefore, namenodes are required to be deployed using on-demand VMs.
The number of namenodes is typically small compared to the number of datanodes (orders of magnitudes) so this will only marginally increase cost and keeps the design simple.
Additionally, imbalances in the usage of datanodes can build up over time depending on the workloads and lifetime of data.
We currently do not have any mechanisms to rebalance the datastore.

\begin{figure*}
  \centering
  \begin{minipage}{0.45\textwidth}
    \centering
    \includegraphics[width=0.85\linewidth]{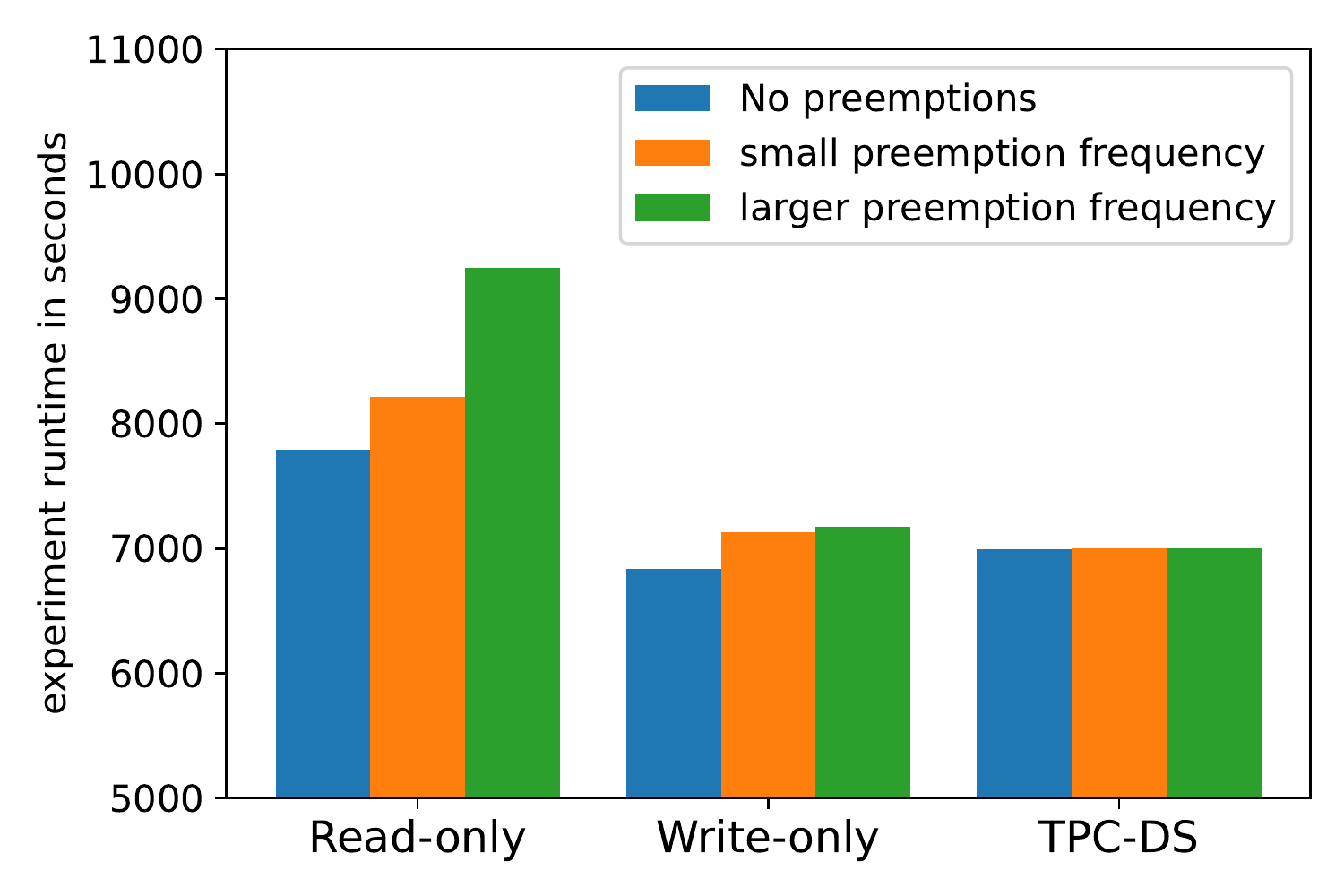}
    \captionof{figure}{Experiment runtimes on local K8s cluster.}
    \label{fig:results_local}
  \end{minipage}%
  \hspace{0.05\textwidth}
  \begin{minipage}{0.45\textwidth}
    \centering
    \includegraphics[width=0.85\linewidth]{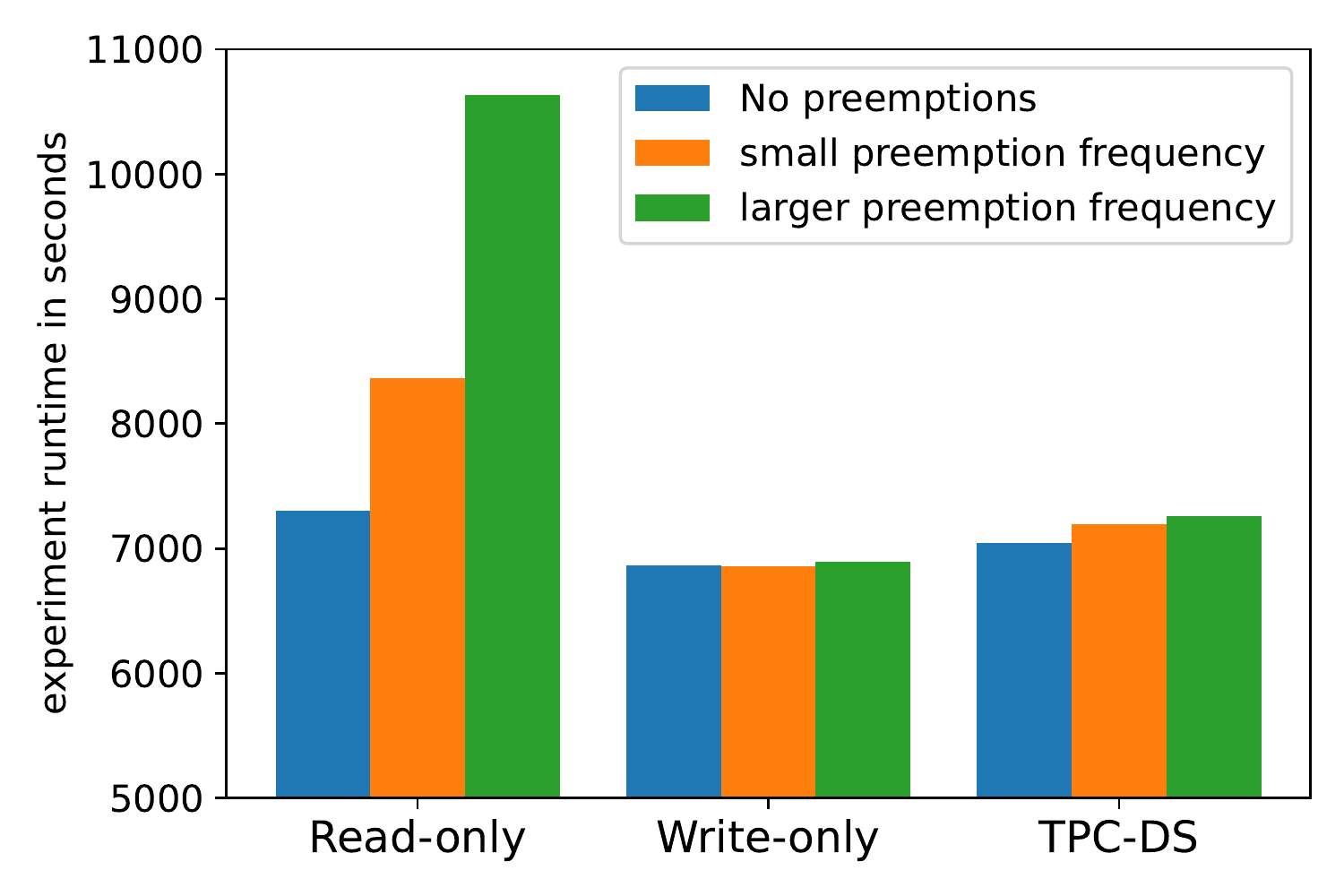}
    \captionof{figure}{Experiment runtimes on GKE K8s cluster.}
    \label{fig:results_gke}
  \end{minipage}
\end{figure*}
\section{Experimental Evaluation}
\subsection{Experiment Methodology}
\textbf{Cluster setup:} We evaluate our system on Google Cloud n2-standard spot instances of various sizes shown in Table~\ref{tab:specs_gke} with an 8-node GKE cluster. We also simulate preemptions in experiments on a local cluster of 8 nodes with specifications shown in Table~\ref{tab:specs_local}. We use DRAM on nodes to store the ephemeral data blocks, since we target applications that require high-bandwidth access to ephemeral data. However, our techniques are applicable to other storage tiers that Crail supports, such as SSD and disk tiers. We deploy datastore components and client applications using Kubernetes. On both setups, we deploy one instance of the Crail namenode and the Relocator process, which run on a on-demand VM rather than a spot instance.\newline
\begin{table}
  \centering
  \begin{tabular}{ | c | c | } 
    \hline
    CPU &  16vCPUs Intel(R) Cascade Lake\\ 
    \hline
    RAM & 64GB \\ 
    \hline
    Network & up to 32 Gbit/s (egress) \\ 
    \hline
  \end{tabular}
  \caption{GKE cluster n2-standard-16 VM specification.}
  \label{tab:specs_gke}
\end{table}
\begin{table}
  \centering
  \begin{tabular}{ | c | c | } 
    \hline
    CPU &  $2\times$ 8-core Intel(R) Xeon(R) E5-2690\\ 
    \hline
    RAM & 94GB \\ 
    \hline
    NIC & Mellanox ConnectX-4 100GbE\\ 
    \hline
  \end{tabular}
  \caption{Local cluster machine specification.}
  \label{tab:specs_local}
\end{table}
\textbf{Modeling spot instance preemptions:} Since we found that the n2-standard spot instances we used in our experiments did not experience high failure rates (we would often need to run for over 24 hours to get even a single spot instance preemption), we also experimented with emulating preemptions based on various frequency distributions. We use a model derived by analyzing the preemption frequency of various spot instance types in Google Cloud from prior work [10]. Figure 2 shows the CDFs of instance preemptions when using the model preemption parameters for two traces that correspond to preemption frequencies for 16-vCPU instances and 32-vCPU instances. In general, larger instance types are more likely to be preempted. \newline
\textbf{Workloads}: We implemented two throughput sensitive micro-benchmark workloads. Our read-only microbenchmark operates on a static long-lived dataset and issues  a series of read operations using parallel threads while a write-only workload writes a set of short-lived data to our datastore. Finally, in our third workload we run a sequence of TPC-DS queries using Apache Spark as an example of a modern data analytics workload and use our datastore for storing all generated ephemeral data. All experiments are designed to take approximately 2 hours each.
\subsection{Datastore Performance}
For each workload, we compare the total execution time when preemptions are generated using the model described above against a baseline without preemptions.
The results are shown in Figure~\ref{fig:results_local} and Figure~\ref{fig:results_gke}.
The execution time of the read-only workload increases significantly when introducing preemptions, on both clusters. This is due to data imbalance across datanodes. Some instances are preempted and restarted early, while others are running longer without being interrupted. Long running datanode instances store more data than short running ones as they receive more data with every new data redistribution. These imbalance leads to unequal load across the available datanodes, and since in its current implementation Crail is not able to perform re-balancing, they result in the observed performance loss.\newline
However, there is minimal performance impact for write-heavy and the TPC-DS workloads. For instance, on the GKE cluster the TPC-DS workload experienced a slowdown of only 2.1\% and 3.0\% for the smaller and large preemption frequencies, respectively. In the write-only and TPC-DS workloads, data is short-lived, in contrast to the read workload. Therefore, no substantial imbalances are built up across the datanodes. The small performance penalty observed here is only caused by the bandwidth overhead of redistributing data after preemptions and the reduced datastore bandwidth during the amount of time it takes to replace the preempted datanode.

\begin{figure}%
  \centering%
  \includegraphics[width=0.80\linewidth]{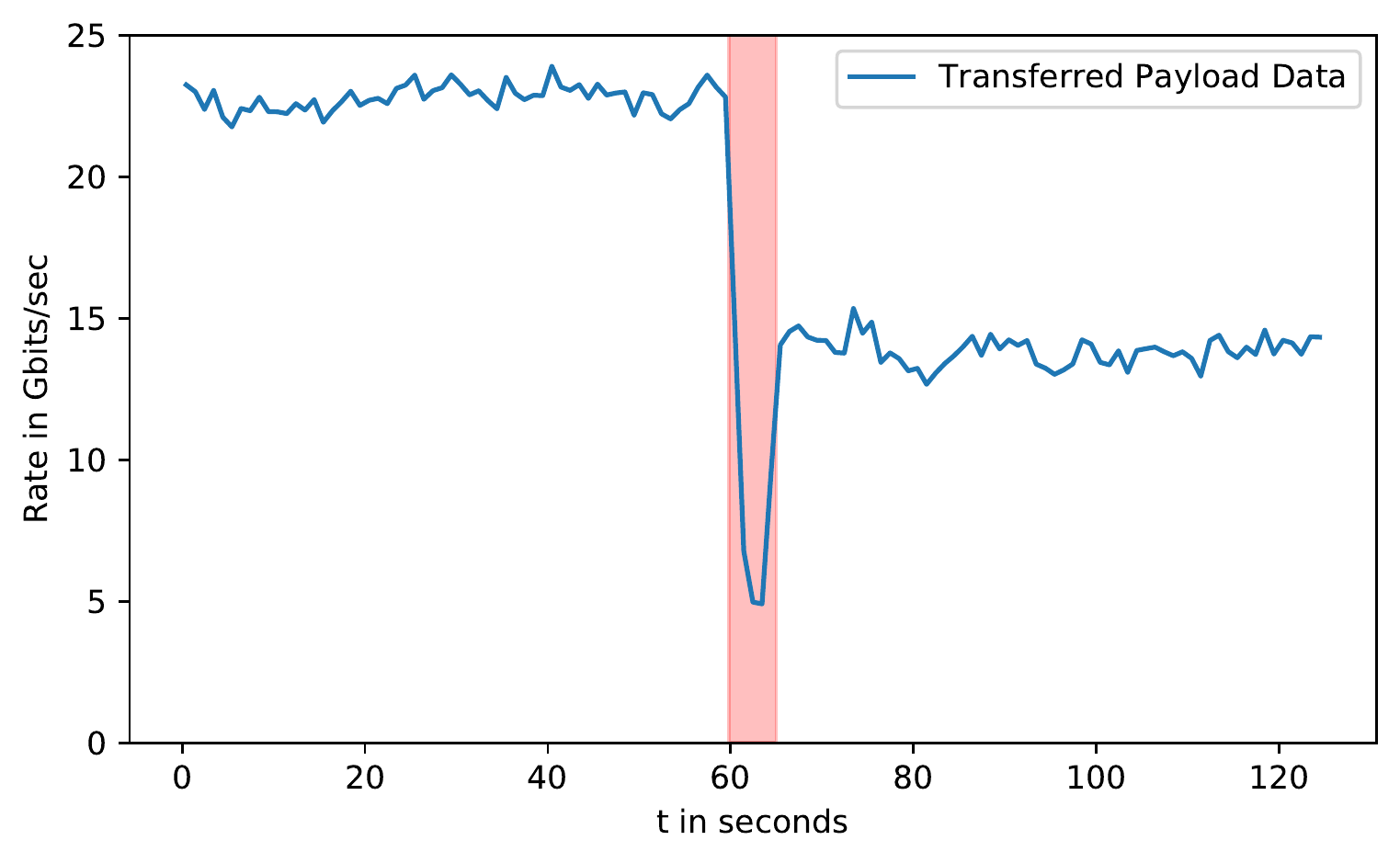}%
  \captionof{figure}{Datastore bandwidth over time after triggering the data redistribution mechanism and terminating one of two instances. Within the area marked in red the data redistribution takes place.}%
  \label{fig:data_migration}%
\end{figure}%

\subsection{Analysis of data migration}
In Figure~\ref{fig:data_migration} we show the impact of migrating data during relocation and of a reduced size datanode cluster before replacing the preempted datanode, by analyzing the bandwidth measured by a read-only client application over time. At the beginning of the experiment two datanodes serve client requests. After sending a preemption signal to one of the datanodes the observed bandwidth drops due to the overhead of redistributing the data stored on the affected instance. Afterwards, the rate stabilizes at a lower level compared to before as the number of datanodes serving requests was reduced. These short bandwidth drops are also observable during the large-scale experiments discussed earlier and cause the performance drops with respect to the baseline. However, since the gaps between preemptions are sufficiently large, on average, the bottom line effect on performance is small.

\subsection{Cost savings}
\begin{figure}%
  \centering%
  \includegraphics[width=0.95\linewidth]{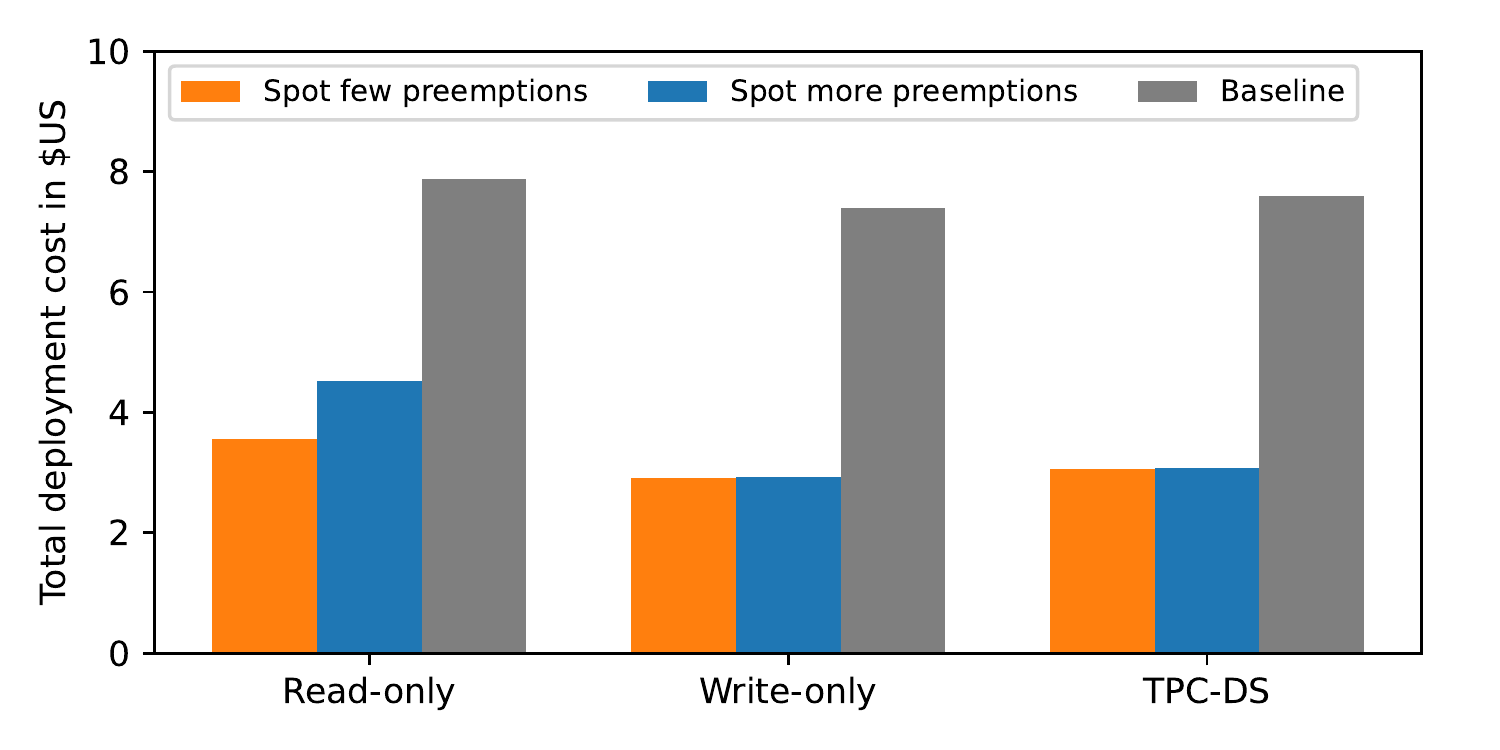}%
  \captionof{figure}{Cost of running a datastore deployment for the entire experiment duration. The baseline uses five on-demand instances and the spot instance deployment uses 4 spot instances and one on-demand instance. When conducting experiments a single on-demand instance cost \$0.776944 and a single spot-instance \$0.188320 for the n2-standard-16 instance type. Spot instance prices are updated at most once per month in the Google Cloud.}%
  \label{fig:cost_comparison}%
\end{figure}%
In Figure~\ref{fig:cost_comparison} we compare the total costs of running a
data\-store deployment consisting of five instances for the entire experiment duration in the Google cloud. When using our proposed data migration mechanism, four instances can be configured as spot instances (one on-demand instance is still required for the metadata node). The baseline deployment, however, uses regular on-demand instances. Due to the significantly lower prices of spot instances, the spot instance deployment achieves approximately 60\% cost savings compared to using dedicated instances, despite the slightly higher runtime.

\subsection{Datanode Capacity Sizing}
\begin{figure}[t!]%
    \centering%
    \includegraphics[width=0.90\linewidth]{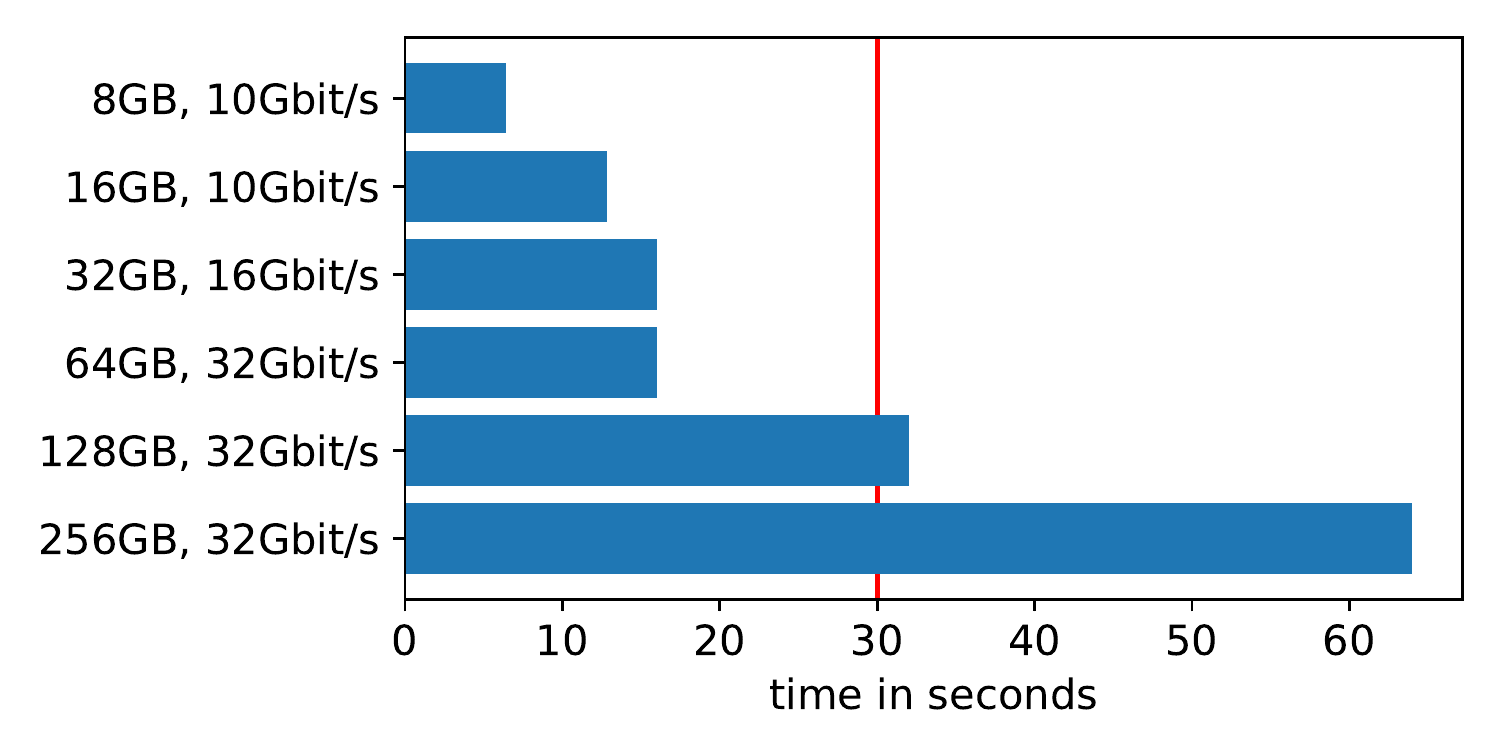}%
    \captionof{figure}{Time required for different sizes of the n2-standard instace type for redistributing the entire memory at the maximum egress bandwidth.}%
    \label{fig:relocation_time}%
  \end{figure}

Due to limited time and other factors like the amount of client traffic reaching an instance and the available network bandwidth the success of data migration cannot be guaranteed.
In Figure~\ref{fig:relocation_time} we show the time required to transfer the amount of data that would fit into the main memory of an instance at the maximum available egress bandwidth for different size configurations of the Google Cloud n2-standard instance type. For larger instance configurations migrating all stored data within the time window is more difficult as the available egress bandwidth does not scale linearly with the memory capacity. Since the prices of instances typically scale linearly with the number of vCPUs and the amount of memory, using several smaller instances would increase the chance of relocating all data in time without increasing costs. Additionally, small instance types are reported to be preempted less frequently compared to large instance types\cite{kadupitige2020modeling}.

\section{Related Work}
\textbf{Spot Instance Systems} An extensively studied application of spot instances for reducing the cost of running workloads in the cloud is to use them for running batch processing jobs \cite{sharma2016flint,menache2014demand,yi2010reducing,song2012optimal}. In this context failed jobs can usually be restarted and systems employ \emph{checkpointing} where progress is periodically saved to remote persistent storage. A lot of attention was also given to adapting fault-tolerant distributed data processing engines like Apache Spark or Hadoop for running processing tasks on top of spot instances \cite{ambati2020providing,chohan2010see,yan2016tr}. In case of spot instance preemptions the systems reschedule affected computations to other available machines in a cluster. Similarly to our proposed system previous work also studied how to harvest the memory resources of spot instances. Examples hereof are in-memory caching layers running on top of instances \cite{wang2017exploiting} and databases serving read requests to frequently accessed items using spot instances \cite{xu2016blending}. \newline
\textbf{Distributed Filesystems} Misra et al. \cite{misra2017scaling} presented a data migration mechanism for moving data between HDFS clusters. While they use data migration to balance load across clusters, we leverage data migration to minimize data loss in case of spot instance preemptions.  Zhang et al. \cite{zhang2016history} face the challenge of preventing data loss and ensuring availibility by using intelligent data replication mechanisms. In Crail replication is not used (mostly for performance reasons).


\section{Conclusion}
We showed how to leverage spot instances to run a stateful ephemeral datastore service at 60\% lower cost. Our data redistribution mechanism minimizes data loss by using the preemption notice period to migrate as much data as possible across other available nodes in the cluster, transparent to application clients. Our evaluation showed that we can achieve these 60\% cost savings with only a 2\% slowdown for a representative data processing workload (TPC-DS).
\newpage

\bibliographystyle{ACM-Reference-Format}
\bibliography{ms}
\end{document}